\documentclass[aps,floats]{revtex4}
\usepackage{amsmath,amssymb}
\pdfoutput=1
\usepackage{graphicx,epsfig}
\usepackage[greek,english]{babel}
\begin{document}
\bibliographystyle {plain}

\def\oppropto{\mathop{\propto}} 
\def\opsimeq{\mathop{\simeq}}
\def\opoverderline{\mathop{\overline}}
\def\operarrow{\mathop{\longrightarrow}}
\def\opsim{\mathop{\sim}}

\def\fig#1#2{\includegraphics[height=#1]{#2}}
\def\figx#1#2{\includegraphics[width=#1]{#2}}

%\newcommand{\fig}[2]{\epsfxsize=#1\epsfbox{#2}} \reversemarginpar 

%%%%%%%%%%%%%%%%%%%%%%%%%%%%%%%%%%%%%%%%%%%%%%%%%%%%%%%%%%%%%%%%%%%%%%%%%%%%
\title{  Random Transverse Field Spin-Glass Model on the Cayley tree : \\
phase transition between the two Many-Body-Localized Phases } 

%%%%%%%%%%%%%%%%%%%%%%%%%%%%%%%%%%%%%%%%%%%%%%%%%%%%%%%%%%%%%%%%%%%%%%%%%%%%

\author{ C\'ecile Monthus }
 \affiliation{Institut de Physique Th\'{e}orique, 
Universit\'e Paris Saclay, CNRS, CEA,
91191 Gif-sur-Yvette, France}

\begin{abstract}
The quantum Ising model with random couplings and random transverse fields on the Cayley tree is studied by Real-Space-Renormalization in order to construct the whole set of eigenstates. The renormalization rules are analyzed via large deviations. The phase transition between the paramagnetic and the spin-glass Many-Body-Localized phases involves the activated exponent $\psi=1$ and the correlation length exponent $\nu=1$. The spin-glass-ordered cluster containing $N_{SG}$ spins is found to be extremely sparse with respect to the total number $N$ of spins : its size grows only logarithmically at the critical point $N_{SG}^{criti} \propto \ln N$, and it is sub-extensive $N_{SG} \propto N^{\theta}$ in the finite region of the spin-glass phase where the continuously varying exponent $\theta$ remains in the interval $0<\theta<1$.

\end{abstract}

\maketitle

\section{ Introduction }

In the field of Many-Body-Localization (see the recent reviews \cite{revue_huse,revue_altman,revue_vasseur,revue_imbrie,revue_rademaker,review_mblergo,review_prelovsek,review_rare,review_toulouse} and references therein), one of the important characterization of Many-Body-Localized phases is the existence
 of an extensive number of Local Integrals of Motion called LIOMs
\cite{emergent_swingle,emergent_serbyn,emergent_huse,emergent_ent,imbrie,serbyn_quench,emergent_vidal,emergent_ros,
emergent_rademaker,serbyn_powerlawent,c_emergent,ros_remanent,c_liom}. 
Since these LIOMS are the building blocks of the whole set of eigenstates, 
it is natural to try to identify them via some real-space renormalization procedure.
The Strong Disorder Real-Space RG approach 
developed by Daniel Fisher \cite{fisher_AF,fisher,fisherreview} to construct the ground states of random quantum models (see the review \cite{strong_review}) has been thus generalized into the RSRG-X procedure to construct the whole set of excited eigenstates \cite{rsrgx,rsrgx_moore,vasseur_rsrgx,yang_rsrgx,rsrgx_bifurcation} : the idea is that each local renormalization step produces a LIOM that describes the choice between the local energy levels (instead of projecting always onto the lowest energy-level).
The RSRG-t procedure developed by Vosk and Altman \cite{vosk_dyn1,vosk_dyn2}
in order to construct the effective dynamics via the iterative elimination of the degree of freedom oscillating with the highest local eigenfrequency is equivalent to the RSRG-X procedure but gives an interesting different point of view \cite{c_rsrgt}.

Since the purpose of these Strong Disorder RG procedures is to produce an extensive number of LIOMS,
it is clear that their validity is limited to Many-Body-Localized Phases : they cannot be applied in the delocalized ergodic phase,
and they do not allow to analyze the MBL transition towards this delocalized phase.
In particular, it should be stressed that the current RG descriptions of the MBL delocalization transition 
 are based on completely different RG rules concerning the entanglement \cite {vosk_rgentanglement},
  the resonances \cite{vasseur_resonant,vasseur_resonantbis}, or the decomposition into insulating and thermal blocks \cite{huse_rgpaste}.
However, the RSRG-X is very useful in MB-Localized phases to analyse the long-ranged order of the excited eigenstates made of LIOMs
and the possible phase transitions between different Many-Body-Localized phases, as for instance the 
transition between the paramagnetic and the spin-glass Many-Body-Localized phases for the one-dimensional generalized quantum Ising model \cite{rsrgx}.

In the present paper, we wish similarly to analyse the transition between the paramagnetic and the spin-glass Many-Body-Localized phases
for the quantum Ising model with random couplings and random transverse fields on the Cayley tree.
Since the standard RSRG-X procedure destroys the tree structure and could only be followed numerically,
we will instead use an RG procedure that preserves the tree structure in order to obtain some simple analytical approximation :
the Pacheco-Fernandez block-RG introduced for the ground state of the one-dimensional chain without disorder
 \cite{pacheco,igloiSD} or with disorder \cite{nishiRandom,us_pacheco,us_renyi} is applied here sequentially 
\cite{us_watermelon} around the center of the tree in order to construct the whole set of eigenstates.

The paper is organized as follows.
In section \ref{sec_rsrgx}, the real-space RG procedure to construct the set of eigenstates
of the random quantum Ising model on the Cayley tree is described.
Section \ref{sec_pr} is devoted to the large deviation properties of the basic variables that appear in the RG flows.
The statistics of the renormalized couplings and of the renormalized transverse field of the center
are studied in section \ref{sec_jr} and in section \ref{sec_hr} respectively in order to characterize the critical properties
of the transition between the paramagnetic and spin-glass Many-Body-Localized phases.
Our conclusions are summarized in section \ref{sec_conclusion}.

\section{ Real-Space RG procedure to construct the set of eigenstates  }

\label{sec_rsrgx}

\subsection{ Model}

We consider the geometry of a Cayley tree of branching ratio $K$ with $L$ generations around the central spin $\sigma_0$.
It is convenient to decompose the quantum Ising Hamiltonian in terms of the contributions
of the various generations
\begin{eqnarray}
H && = \sum_{r=0}^L H_r
\nonumber \\
H_0 && =  h_0 \sigma_0^z 
 \nonumber \\
H_1 && =  \sum_{i_1=1}^{K+1} ( J_{i_1} \sigma_0^x \sigma_{i_1}^x  + h_{i_1} \sigma_{i_1}^z )
 \nonumber \\
H_2 && = \sum_{i_1=1}^{K+1} \sum_{i_2=1}^{K}  ( J_{i_1,i_2} \sigma_{i_1}^x \sigma_{i_1,i_2}^x+ h_{i_1,i_2} \sigma_{i_1,i_2}^z) 
\nonumber \\  
 H_r && = \sum_{i_1=1}^{K+1} \sum_{i_2=1}^{K} .. \sum_{i_r=1}^{K}
( J_{i_1,i_2,..,i_r} \sigma_{i_1,..,i_{r-1}}^x \sigma_{i_1,..,i_r}^x   
 + h_{i_1,..,i_r} \sigma_{i_1,..,i_r}^z )
\label{htree}
\end{eqnarray}
We consider 
that both the transverse fields $h_{i_1,i_2,..,i_r}$ and the couplings $J_{i_1,i_2,..,i_r}$ 
are random variables drawn with some continuous distributions.
As example, we will focus on the case where the probability distributions of the couplings $J_{i_1,i_2,..,i_r}$ and of the random fields $h_{i_1,i_2,..,i_r}$ are uniformly drawn on $[-J,+J]$ and $[-h,+h]$ respectively 
\begin{eqnarray}
\pi_{J}(J_{i_1,i_2,..,i_r}) =\frac{\theta( -J \leq J_{i_1,i_2,..,i_r} \leq J) }{ 2 J} 
\nonumber \\
\pi_{h}(h_{i_1,i_2,..,i_r}) =\frac{\theta( -h \leq h_{i_1,i_2,..,i_r} \leq h) }{ 2 h} 
\label{box}
\end{eqnarray}

This model is MB-Localized in the two following limits :

(i) When all couplings vanish $J_{i_1,i_2,..,i_r} \to 0$, the model is in a trivial Paramagnetic MB-Localized phase,
where the LIOMs (commuting with themselves and the Hamiltonian)
are the site operators $\sigma^z_{i_1,i_2,..,i_r}$ that are coupled to the random fileds $h_{i_1,..,i_r} $
\begin{eqnarray}
\tau^z_{i_1,i_2,..,i_r} \opsimeq_{J_{\{ \}} \to 0}   \sigma^z_{i_1,i_2,..,i_r}
\label{liomjzero}
\end{eqnarray}
Since all the random fields $h_{i_1,..,i_r} $ are different, there is no degeneracy between the many-body-energy-levels,
 and the LIOMs of Eq. \ref{liomjzero} are perturbatively stable
in the presence of small couplings $J_{\{ \}}$ .
Note the difference with the model studied in Ref. \cite{mossi} where the transverse fields all take the same value $h$,
so that the zero-coupling model is the pure paramagnetic model characterized by huge degeneracies in many-body-energy-levels.

(ii) In the opposite limit where all fields vanish $h_{i_1,i_2,..,i_r} \to 0$, the model is in a trivial Spin-Glass Localized phase,
where the LIOMs are the bond operators $\sigma_{i_1,..,i_{r-1}}^x \sigma_{i_1,..,i_r}^x$ associated to the random couplings $J_{i_1,..,i_r} $
\begin{eqnarray}
\tau^z_{i_1,i_2,..,i_r} \opsimeq_{h_{\{ \}} \to 0}   \sigma_{i_1,..,i_{r-1}}^x \sigma_{i_1,..,i_r}^x
\label{liomhzero}
\end{eqnarray}
Since all the couplings $ J_{i_1,..,i_r}$ are different, the many-body-energy-levels are non-degenerate,
 and the LIOMs of Eq. \ref{liomhzero} are perturbatively stable in the presence of small fields $h_{\{ \}} $.
Note again the difference with the model studied in Ref. \cite{mossi} where the couplings only take the two values $(\pm J)$,
leading to huge degeneracies in energy many-body-energy-levels.

In summary, the continuous distributions of both random fields and couplings is necessary to avoid degeneracies between many-body-energy-levels and
to identify simple LIOMs in the two limits of vanishing couplings (Eq \ref{liomjzero}) or vanishing fields (Eq. \ref{liomhzero}).
Since these two type of LIOMs correspond to different Long-Ranged order for the corresponding eigenstates, namely Paramagnetic and Spin-Glass,
one expects that the full model containing both fields and couplings will display
 a phase transition between two different Many-Body-Localized phases (Paramagnetic and Spin-Glass).
The goal of the present paper is to analyse this transition 
via some real-space procedure that constructs the LIOMs and thus the set of eigenstates.

\subsection { First RG step }

The RSRG-X procedure mentioned in the Introduction can be applied in $d>1$, but the changes of the geometry
prevents the finding of any analytical description. The renormalization procedure has to be implemented numerically,
as was done for the RSRG procedure concerning the ground state in $d=2,3,4$ \cite{fisherreview,motrunich,lin,karevski,lin07,yu,kovacsstrip,kovacs2d,kovacs3d,kovacsentropy,kovacsreview}.
Here we wish instead to obtain some analytically solvable RG procedure in order to get more insight
into the mechanism of the transition.
We have thus chosen to apply sequentially \cite{us_watermelon} around the center of the tree
the idea of the Pacheco-Fernandez elementary step  \cite{pacheco,igloiSD,nishiRandom,us_pacheco,us_renyi}
in order to keep a simple geometry along the RG flow.

More precisely, the first RG step consists in the diagonalization
of the Hamiltonian $H_1$ Eq. \ref{htree} concerning the center spin and the $(K+1)$ spins 
of the first generation
\begin{eqnarray}
H_{1} =
  \sum_{i_1=1}^{K+1} \left( J_{i_1} \sigma_0^x \sigma_{i_1}^x + h_{i_1} \sigma_{i_1}^z   \right) 
\label{h1c}
\end{eqnarray}
Since $H_1$ commutes with $\sigma_0^x$, one needs to consider the two possible values $ \sigma_0^x=S_0^x=\pm 1$,
and to diagonalize the $(K+1)$ remaining effective Hamiltonians involving the single spin $\sigma_{i_1}$ 
\begin{eqnarray}
H^{eff}_{i_1} =  J_{i_1} S_0^x \sigma_{i_1}^x + h_{i_1} \sigma_{i_1}^z  
\label{heffi1}
\end{eqnarray}
The two eigenvalues of Eq. \ref{heffi1} do not depend on the value $S_0^x=\pm$ and read
\begin{eqnarray}
\lambda^{(\tau_{i_1}^z) }_{i_1} = \tau_{i_1}^z \sqrt{  J_{i_1}^2+h_{i_1}^2 } 
\label{lambdai1}
\end{eqnarray}
where the variable
\begin{eqnarray}
\tau_{i_1}^z = \pm
\label{tauiz}
\end{eqnarray}
labels the choice between the positive or negative energy in Eq. \ref{lambdai1}.
The corresponding eigenvectors depend on the value $S_0^x$
\begin{eqnarray}
\vert \lambda^{(\tau_{i_1}^z) }_{i_1}(S_0^x) > = 
\sqrt{ \frac{1}{2} \left( 1+\frac{\tau_{i_1}^z S_0^x J_{i_1} }{\sqrt{ J_{i_1}^2 + h_{i_1}^2 }}  \right) }  \vert \sigma_{i_1}^x=+ 
\rangle
+ \tau_{i_1}^z {\rm sgn}(h_{i_1}) \sqrt{ \frac{1}{2} \left( 1- \frac{\tau_{i_1}^z S_0^x J_{i_1} }{\sqrt{ J_{i_1}^2 + h_{i_1}^2 }}  \right) }  \vert \sigma_{i_1}^x=- \rangle
\label{lambdatauiz}
\end{eqnarray}

To make the link with the Lioms of Eq. \ref{liomjzero} and \ref{liomhzero}, it is usefule to condider the two corresponding limits :

(i) if the coupling vanishes $J_1=0$, the eigenvalues and eigenvectors reduce to
\begin{eqnarray}
\lambda^{(\tau_{i_1}^z) }_{i_1} && \opsimeq_{J_1=0}  \tau_{i_1}^z \vert h_{i_1} \vert  = \tau_{i_1}^z {\rm sgn}(h_{i_1})  h_{i_1}
\nonumber \\
\vert \lambda^{(\tau_{i_1}^z) }_{i_1}(S_0^x) > && \opsimeq_{J_1=0} 
 \frac{\vert \sigma_{i_1}^x=+ \rangle
+ \tau_{i_1}^z {\rm sgn}(h_{i_1})  \vert \sigma_{i_1}^x=- \rangle}{ \sqrt 2  }  = \vert \sigma_{i_1}^z= \tau_{i_1}^z {\rm sgn}(h_{i_1}) \rangle
\label{lambdatauizjzero}
\end{eqnarray}
that is equivalent to Eq. \ref{liomjzero} up to the factor $ {\rm sgn}(h_{i_1})$ that comes from the choice of Eq. \ref{tauiz} to label the sign of the energy of Eq. \ref{lambdai1}.

(ii) if the field vanishes $h_{i_1}=0$,  the eigenvalues and eigenvectors become
\begin{eqnarray}
\lambda^{(\tau_{i_1}^z) }_{i_1} && \opsimeq_{h_1=0}  \tau_{i_1}^z \vert J_{i_1} \vert  = \tau_{i_1}^z {\rm sgn}(J_{i_1})  J_{i_1}
\\
\vert \lambda^{(\tau_{i_1}^z) }_{i_1}(S_0^x) > &&  \opsimeq_{h_1=0} 
\sqrt{ \frac{1+\tau_{i_1}^z S_0^x {\rm sgn} ( J_{i_1})}{2}  }  \vert \sigma_{i_1}^x=+  \rangle
+ \tau_{i_1}^z {\rm sgn}(h_{i_1}) 
\sqrt{ \frac{1-\tau_{i_1}^z S_0^x {\rm sgn} ( J_{i_1})}{2}  }   \vert \sigma_{i_1}^x=- \rangle
\propto  \vert \sigma_{i_1}^x=\tau_{i_1}^z S_0^x {\rm sgn} ( J_{i_1})  \rangle
\nonumber 
\label{lambdatauizhzero}
\end{eqnarray}
that is equivalent to Eq. \ref{liomhzero} up to the factor $ {\rm sgn}(J_{i_1})$ that comes from the choice of Eq. \ref{tauiz} to label the sign of the energy of Eq. \ref{lambdai1}.

When the coupling $J_{i_1}$ and the field $h_{i_1}$ are both non-vanishing,
the LIOM $\tau_{i_1}^z$ defined by Eqs \ref{lambdai1} and \ref{lambdatauiz} can be thus considered as the appropriate 
interpolation between these two simple limits (i) and (ii).
Note that in usual Strong-Disorder RG rules for MB-Localized phases \cite{rsrgx}, each LIOM is declared to be associated either to a site variable as in (i)
 (if its renormalized transverse field is the biggest among surviving variables) or to a bond variable as in (ii) 
  (if its renormalized coupling is the biggest among surviving variables), so that each LIOM could be called accordingly 'paramagnetic' or 'spin-glass'.
On the contrary, within the present procedure, the LIOM $\tau_{i_1}^z$ is some interpolation between (i) and (ii) 
as in the block-RG procedures of \cite{c_emergent}, and thus cannot be called 'paramagnetic' or 'spin-glass' in itself.

Let us now return to the whole Hamiltonian $H_1$ of Eq. \ref{h1c} : the $2^{K+1}$ energy-levels labelled by the variables $(\tau_1^z,...\tau_{K+1}^z )$
\begin{eqnarray}
E_{1}^{(\tau_1^z...\tau_{K+1}^z ) } = \sum_{i_1=1}^{K+1} \tau_{i_1}^z \sqrt{ J_{i_1}^2 + h_{i_1}^2 }
\label{e1level}
\end{eqnarray}
are independent of $S_0^x=\pm 1$.
To label this degeneracy, it is thus convenient to introduce the renormalized spin $\sigma_{R0}^x$
\begin{eqnarray}
\vert \tau_1^z...\tau_{K+1}^z ; \sigma_{R0}^x=+1\rangle  && =\vert  S_{0}^x=+1 \rangle
   \otimes_{i_1=1}^{K+1} \vert \lambda^{(\tau_{i_1}^z) }_{i_1}(S_0^x=+1) \rangle
\nonumber \\
\vert \tau_1^z...\tau_{K+1}^z ; \sigma_{R0}^x=-1  \rangle  && =\vert  S_{0}^x=-1 \rangle
   \otimes_{i_1=1}^{K+1} \vert \lambda^{(\tau_{i_1}^z) }_{i_1}(S_0^x=-1) \rangle
\label{tausr}
\end{eqnarray}
The projector onto the energy-level $ E_{1}^{(\tau_1^z...\tau_{K+1}^z ) }$ then reads
\begin{eqnarray}
P_{1}^{(\tau_1^z...\tau_{K+1}^z ) } =
 \vert \tau_1^z...\tau_{K+1}^z ; \sigma_{R0}^x=+1\rangle \langle \tau_1^z...\tau_{K+1}^z ; \sigma_{R0}^x=+1 \vert
+  \vert \tau_1^z...\tau_{K+1}^z ; \sigma_{R0}^x=-1  \rangle \langle \tau_1^z...\tau_{K+1}^z ; \sigma_{R0}^x=-1 \vert
\label{projector}
\end{eqnarray}

The projection onto the energy-level $ E_{1}^{(\tau_1^z...\tau_{K+1}^z ) }$
of the Hamiltonian of Eq \ref{htree} concerning the whole tree can be obtained from the various contributions
\begin{eqnarray}
P_{1}^{(\tau_1^z...\tau_{K+1}^z ) } H P_{1}^{(\tau_1^z...\tau_{K+1}^z ) }&& = \sum_{r=0}^L P_{1}^{(\tau_1^z...\tau_{K+1}^z ) } H_r P_{1}^{(\tau_1^z...\tau_{K+1}^z ) } 
\label{htreeproj}
\end{eqnarray}
The projection of $H_1$ is simply the energy $ E_{1}^{(\tau_1^z...\tau_{K+1}^z ) }$ by construction
\begin{eqnarray}
P_{1}^{(\tau_1^z...\tau_{K+1}^z ) } H_1 P_{1}^{(\tau_1^z...\tau_{K+1}^z ) } && =   E_{1}^{(\tau_1^z...\tau_{K+1}^z ) }
\label{h1treeproj}
\end{eqnarray}
while the projection of $H_{r }$ is unchanged for $r \geq 3$
\begin{eqnarray}
P_{1}^{(\tau_1^z...\tau_{K+1}^z ) } H_{r } P_{1}^{(\tau_1^z...\tau_{K+1}^z ) }&& = H_r
\label{hrtree}
\end{eqnarray}

The projection of $H_0$
\begin{eqnarray}
P_{1}^{(\tau_1^z...\tau_{K+1}^z ) } H_0 P_{1}^{(\tau_1^z...\tau_{K+1}^z ) } && =  h_0 P_{1}^{(\tau_1^z...\tau_{K+1}^z ) } \sigma_0^z P_{1}^{(\tau_1^z...\tau_{K+1}^z) } 
\nonumber \\
&& = h_0
\left( \prod_{i_1=1}^{K+1}  \sqrt{\frac{h_{i_1}^2 }{ J_{i_1}^2 + h_{i_1}^2}  } \right) 
P_{1}^{(\tau_1^z...\tau_{K+1}^z ) }  \sigma_{R0}^z P_{1}^{(\tau_1^z...\tau_{K+1}^z ) } 
\label{h0tproj}
\end{eqnarray}
gives the renormalized transverse field $h^R_{0} $ associated to the renormalized spin operator $\sigma_{R0}^z $
\begin{eqnarray}
h^R_{0}  = h_0 \prod_{i_1=1}^{K+1}   \sqrt{\frac{h_{i_1}^2 }{ J_{i_1}^2 + h_{i_1}^2}  } 
\label{hr0}
\end{eqnarray}

The projection of $H_2$
\begin{eqnarray}
P_{1}^{(\tau_1^z...\tau_{K+1}^z ) }H_2 P_{1}^{(\tau_1^z...\tau_{K+1}^z ) }
&& = \sum_{i_1=1}^{K+1} \sum_{i_2=1}^{K}  \left( J_{i_1,i_2} 
(P_{1}^{(\tau_1^z...\tau_{K+1}^z ) } \sigma_{i_1}^x P_{1}^{(\tau_1^z...\tau_{K+1}^z ) }) \sigma_{i_1,i_2}^x+ h_{i_1,i_2} \sigma_{i_1,i_2}^z
\right) 
\nonumber \\
&& = \sum_{i_1=1}^{K+1} \sum_{i_2=1}^{K}  \left( J_{i_1,i_2} \frac{\tau_{i_1}^z J_{i_1} }{\sqrt{ J_{i_1}^2 + h_{i_1}^2 }}
(P_{1}^{(\tau_1^z...\tau_{K+1}^z ) } \sigma_{R0}^x P_{1}^{(\tau_1^z...\tau_{K+1}^z ) }) \sigma_{i_1,i_2}^x+ h_{i_1,i_2} \sigma_{i_1,i_2}^z
\right) 
\label{h2proj}
\end{eqnarray}
gives the renormalized coupling between the operators $\sigma_{R0}^x $ and $ \sigma_{i_1,i_2}^x$ 
\begin{eqnarray}
J_{i_1,i_2}^R =  J_{i_1,i_2}  \frac{\tau_{i_1}^z  J_{i_1} }{ \sqrt{ J_{i_1}^2 + h_{i_1}^2 } }
\label{jr12}
\end{eqnarray}

\subsection { RG rules }

The iteration of the above procedure yields the following RG rules after $r$ RG steps.
The renormalized transverse field $h^{R^r}_{0} $ associated to the renormalized spin operator $\sigma_{R^r0}^z $ 
evolves
according to
(Eq \ref{hr0}) 
\begin{eqnarray}
h_0^{R^r} = h_0^{R^{n-1}} \prod_{i_1=1}^{K+1} \prod_{i_2=1}^{K}  .. \prod_{i_r=1}^{K} 
\sqrt{  \frac{ h_{i_1,..,i_r}^2 }
{  h_{i_1,..,i_r}^2 + [ J_{i_1,..,i_r}^{R^{r-1}} ]^2  } }
\label{rgh}
\end{eqnarray}
while the renormalized coupling between the operators $\sigma_{R^r0}^x $ and $ \sigma_{i_1,i_2,..,i_{r+1}}^x$ reads (Eq \ref{jr12})
\begin{eqnarray}
J_{i_1,..,i_{r+1}}^{R^r} =  J_{i_1,...,i_{r+1}} \frac{\tau_{i_1,..,i_r}^z  J_{i_1,..,i_{r}}^{R^{r-1}} }
{ \sqrt{  h_{i_1,..,i_r}^2 + [ J_{i_1,..,i_r}^{R^{r-1}} ]^2  } }
\label{rgj}
\end{eqnarray}

\subsection { Solution of the RG rules }

The RG rule of Eq. \ref{rgj} for the couplings involve only the initial transverse fields 
and not the renormalized transversed fields, so that it can be solved independently.
The sign
\begin{eqnarray}
{\rm sgn}( J_{i_1,..,i_{r+1}}^{R^r} ) && = \tau_{i_1,..,i_r}^z {\rm sgn}( J_{i_1,...,i_{r+1}} ) {\rm sgn} (  J_{i_1,..,i_{r}}^{R^{r-1}} )
\nonumber \\
&& = \tau_{i_1,..,i_r}^z \tau_{i_1,..,i_{r-1}}^z ... \tau_{i_1}^z 
{\rm sgn}( J_{i_1,...,i_{r+1}} )  {\rm sgn}( J_{i_1,...,i_{r}} ) ... {\rm sgn}( J_{i_1} ) 
\label{rgjsgn}
\end{eqnarray}
is simply the product of all the couplings $ J$ and of all the variables$\tau^z$ along the path between 
the sites $0$ and $(i_1,..,i_r)$.

The absolute value reads (Eq. \ref{rgj})
\begin{eqnarray}
\vert J_{i_1,..,i_{r+1}}^{R^r} \vert && =\vert  J_{i_1,...,i_{r+1}} \vert  C_{i_1,..,i_r}
\label{rgjabs}
\end{eqnarray}
where
\begin{eqnarray}
C_{i_1,..,i_r} && \equiv \left[ 1+\sum_{m=1}^r \prod_{k=m}^r \frac{h_{i_1,..,i_k}^2 }{J_{i_1,..,i_k}^2} \right]^{-\frac{1}{2}}
\nonumber \\
&& = \left[ 1+\frac{h_{i_1,..,i_r}^2 }{J_{i_1,..,i_r}^2} + \frac{h_{i_1,..,i_r}^2 h_{i_1,..,i_{r-1}}^2 }{J_{i_1,..,i_r}^2 J_{i_1,..,i_{r-1}}^2} + .. +  \frac{h_{i_1,..,i_r}^2 h_{i_1,..,i_{r-1}}^2..h_{i_1,i_2}^2 h_{i_1}^2 }{J_{i_1,..,i_r}^2 J_{i_1,..,i_{r-1}}^2.. J_{i_1,i_2}^2 J_{i_1}^2}
 \right]^{-\frac{1}{2}}
\label{kesten}
\end{eqnarray}
involves in the denominator a so-called Kesten random variable \cite{Kesten,Der_Pom,Bou,Der_Hil,Cal}
that has been much studied in relation with the surface magnetization in the ground-state of the one-dimensional chain
\cite{strong_review,c_microcano,us_watermelon}.

This solution for the renormalized couplings can be plugged into the RG flow of Eq. \ref{rgh}
for the renormalized transverse field to obtain
\begin{eqnarray}
\ln \left( \frac{h_0^{R^r} }{ h_0^{R^{r-1}} } \right) && =  \sum_{i_1=1}^{K+1} \sum_{i_2=1}^{K}  .. \sum_{i_r=1}^{K}
\ln \left(   \frac{ 1 }
{ \sqrt{  1 +\frac{ J^2_{i_1,..,i_r} }{h_{i_1,..,i_r}^2}   C_{i_1,..,i_{r-1}}^2  } }  \right)
\nonumber \\
&& =  \frac{1}{2}   \sum_{i_1=1}^{K+1} \sum_{i_2=1}^{K}  .. \sum_{i_r=1}^{K}
\ln \left( 1- C_{i_1,..,i_{r-1},i_r}^2   \right)
\label{rghbis}
\end{eqnarray}
in terms of the Kesten variables of Eq. \ref{kesten}.

\subsection { Reminder on the one-dimensional chain $K=1$ }

For the one-dimensional chain corresponding to $K=1$, the location of paramagnetic/spin-glass quantum phase transition 
for the ground state of the quantum Ising model is know to occur
exactly at 
\begin{eqnarray}
{\rm Critical \ Point \ in \ one \  dimension : } \ \ \ \overline{\ln \vert J_i \vert } = \overline{\ln \vert h_i \vert}
\label{criti1d}
\end{eqnarray}
 as a consequence of self-duality \cite{pfeuty,fisher,strong_review}.
The corresponding Strong Disorder Fixed Point \cite{fisher} is characterized in particular by the activated exponent
\begin{eqnarray}
\psi^{(d=1)} = \frac{1}{2}
\label{psi1d}
\end{eqnarray}
and by the two correlation length exponents
\begin{eqnarray}
\nu^{(d=1)}_{typ} && =1
\nonumber \\
\nu^{(d=1)}_{av} && = 2
\label{nu1d}
\end{eqnarray}

As discussed in \cite{c_emergent}, the phase transition between the Paramagnetic and Spin-glass
Many-Body-Localized phases for the excited eigenstates is the same as the ground state quantum phase transition just described,
and the above renormalization procedure is able to reproduce the exact transition location of Eq. \ref{criti1d} and the exact critical exponents of 
Eqs \ref{psi1d} and \ref{nu1d}, together with the exact surface magnetization in terms of Kesten variables as already mentioned above (Eq \ref{kesten}).

\subsection { Solution at lowest order in the couplings for the Cayley tree with branching ratio $K>1$}

We have just recalled that in one dimension, the transition occurs when the typical coupling and the typical fields are equal (Eq. \ref{criti1d}).
 For the Cayley tree with branching ratio $K>1$, the transition is thus expected to occur in the region 
\begin{eqnarray}
\overline{\ln \vert J_i \vert } < \overline{\ln \vert h_i \vert}
\label{jtypsmaller}
\end{eqnarray}
where the couplings are typically smaller than the transverse fields.
To analyse the RG rules in this region, it is convenient to introduce the products
\begin{eqnarray}
P_{i_1,..,i_r} && \equiv \left \vert  \frac{J_{i_1,..,i_r} J_{i_1,..,i_{r-1}}.. J_{i_1,i_2} J_{i_1}}{h_{i_1,..,i_r} h_{i_1,..,i_{r-1}}..h_{i_1,i_2} h_{i_1} }
   \right \vert
\label{product}
\end{eqnarray}

In the region of Eq. \ref{jtypsmaller}, the Kesten variable of the denominator in Eq. \ref{kesten} is dominated by the last term,
while it is convenient to keep the term unity to maintain the important bound $C_{i_1,..,i_r } \leq 1$,
so that we make the following approximation at lowest order in the couplings
\begin{eqnarray}
C_{i_1,..,i_r} && \simeq \left[ 1+ \frac{1}{P^2_{i_1,..,i_r}}  \right]^{-\frac{1}{2}}
= \frac{P_{i_1,..,i_r} }{\sqrt{ 1+P^2_{i_1,..,i_r} } }
\label{kestenlow}
\end{eqnarray}
Then the absolute values of the renormalized couplings of Eq \ref{rgjabs} become
\begin{eqnarray}
\vert  J_{i_1,..,i_{r+1}}^{R^r}  \vert && = \vert  J_{i_1,...,i_{r+1}} \vert  \frac{P_{i_1,..,i_r} }{\sqrt{ 1+P^2_{i_1,..,i_r} }}
\label{jrlow}
\end{eqnarray}
For the ground state, the result $\vert  J_{i_1,...,i_{n+1}} \vert  P_{i_1,..,i_r}  $ (i.e. without the denominator $\sqrt{1+P^2_{i_1,..,i_r} }$)
 that involves the product of all couplings in the numerator and all the transverse fields in the denominator 
has been obtained in the paramagnetic phase via various approaches including the Cavity-Mean-Field approach \cite{cavity1,cavity2,cavity3},
the Strong Disorder RG framework when only sites are decimated \cite{us_boundarycayley} or simply 
lowest perturbation theory in the couplings \cite{us_transverseDP}.

The approximation of Eq \ref{kestenlow}
yields that the RG flow of Eq. \ref{rghbis}
for the renormalized transverse field becomes
\begin{eqnarray}
\ln \left( \frac{h_0^{R^r} }{ h_0^{R^{r-1}} } \right) 
&& \simeq  - \frac{1}{2}      \sum_{i_1=1}^{K+1} \sum_{i_2=1}^{K}  .. \sum_{i_r=1}^{K}
\ln \left( 1+ P_{i_1,..,i_{r-1},i_r}^2   \right) 
\label{hrlow}
\end{eqnarray}
To analyse the statistical properties of the RG flows Eq \ref{jrlow} and Eq \ref{hrlow},
one needs first to characterize the large deviation properties of the products of Eq. \ref{product}.

\section{ Large deviation analysis }

\label{sec_pr}

In this section, we describe the statistical properties of the product of Eq. \ref{product}
with the simplified notation 
\begin{eqnarray}
P(r)  &&=   \prod_{k=1}^r \left \vert  \frac{J_{i_1,..,i_k} }{h_{i_1,..,i_k}  }    \right \vert
\label{cradial}
\end{eqnarray}
where $r$ represents the number of random variables $\left \vert  \frac{J_{i_1,..,i_k} }{h_{i_1,..,i_k}  }    \right \vert $ in this product.

\subsection{ Typical behavior }

The logarithm of Eq. \ref{cradial}
reduces to a sum of random variables
\begin{eqnarray}
\ln P(r)  && \simeq  \sum_{k=1}^r (\ln  \vert  J_{i_1,..,i_k} \vert - \ln \vert h_{i_1,..,i_k}   \vert )
\label{lnpr}
\end{eqnarray}
The Central Limit Theorem thus yields the following typical behavior for large $r$ 
\begin{eqnarray}
\ln P(r)  && \opsimeq_{r \to +\infty}   - r a_{0}+ \sqrt{r} u
\label{clt}
\end{eqnarray}
where 
\begin{eqnarray}
a_{0} = \overline{  (\ln  \vert  h_{i} \vert - \ln \vert J_{i}   \vert ) }
\label{a0}
\end{eqnarray}
is positive $a_0>0$ in the region under study (Eq. \ref{jtypsmaller}) and 
governs the typical exponential decay of $P(r)$, while $u$ is a Gaussian random variable.
For the one-dimensional chain, only this typical behavior is relevant,
but here on the Cayley tree of branching ratio $K>1$ 
where the number of sites at distance $r$ grows exponentially as $K^r$ with the distance $r$,
one needs to analyze the large deviations properties.

\subsection{ Large deviations }

In the field of large deviations (see the review \cite{touchette} and references therein),
one is interested into the exponentially small probability 
to see an exponential decay with some coefficient $a $ different from the typical value $a_0$ of Eq. \ref{a0}
\begin{eqnarray}
{\rm Prob} ( P(r) \propto e^{-a r}  ) && \oppropto_{r \to +\infty} e^{-r I(a)} 
\label{largedev}
\end{eqnarray}
where the rate function $I(a)$ vanishes at the typical value $a_0$ (Eq \ref{a0})
\begin{eqnarray}
I(a_0)=0
\label{largedevtyp}
\end{eqnarray}
and is strictly positive otherwise $I(a \ne a_0) >0 $.
The standard way to evaluate the rate function $I(a)$ is to consider
the generalized moments that display the following exponential behavior \cite{touchette} 
\begin{eqnarray}
\overline{ P^{2q} (r)  } && =   \left(  \overline{ \frac{\vert J_i \vert ^{2q} }{\vert h_i \vert ^{2q}  }  }  \right)^r = e^{ r \lambda(q)}
\label{c2q}
\end{eqnarray}
where 
\begin{eqnarray}
 \lambda(q) && =   \ln \left(  \overline{ \frac{\vert J_i \vert ^{2q} }{\vert h_i \vert ^{2q}  }  }  \right)
\label{lambdaq}
\end{eqnarray}
can be explicitly computed from the probability distribution of the couplings $J_i$ and of the random fields $h_i$
(see the example below).
The evaluation of Eq. \ref{c2q} via the saddle-point approximation
\begin{eqnarray}
\overline{ P^{2q} (r)  } && \simeq  \int_0^{+\infty} da e^{-r  I(a)}   e^{-a r 2 q } = e^{r \left( \displaystyle \max_a ( - I(a) - 2q a   ) \right) }
\label{saddle}
\end{eqnarray}
yields $\lambda_q$ in terms of the saddle-point $a_q$
\begin{eqnarray}
\lambda(q) && =  - I(a_q) -  2 q  a_q
\nonumber \\
0 && =   I'(a_q) + 2 q
\label{legendre}
\end{eqnarray}
The reciprocal Legendre transform yields
\begin{eqnarray}
I(a)  && = - \lambda(q_a)   -  2  a q_a
\nonumber \\
0 && =   \lambda'(q_a)   +  2  a
\label{legendrereciproque}
\end{eqnarray}

\subsection{ Explicit example with the two box distributions of Eq. \ref{box} }

Let us now focus on the example where the probability distributions of the couplings and of the random fields
are the two box distributions of parameters $J$ and $h$ respectively (Eq \ref{box}).
In the region $h>J$, the typical decay of the renormalized couplings is governed by (Eq. \ref{a0})
\begin{eqnarray}
a_{0} =\int_0^h \frac{dh_i}{h}  \ln  h_i - \int_0^J \frac{dJ_i}{J} \ln J_i = \ln \left( \frac{h}{J} \right)  >0
\label{a0box}
\end{eqnarray}

The generalized moments of Eq. \ref{c2q}
converge only in the region $-1<2q<1$ and Eq. \ref{lambdaq} becomes
\begin{eqnarray}
 e^{  \lambda(q)} && = \overline{ \frac{\vert J_i \vert ^{2q} }{\vert h_i \vert ^{2q}  }  }  =  \int_0^J \frac{dJ_i}{J} J_i^{2q} \int_0^h \frac{dh_i}{h} h_i^{-2q} 
= \frac{1}{1- 4 q^2} \left(  \frac{J }{h}  \right)^{2q} = \frac{1}{1- 4 q^2} e^{-2q a_0}
\label{lqbox}
\end{eqnarray}
so that the function $\lambda(q)$ and its derivative read in terms of the typical value $a_0$ 
\begin{eqnarray}
 \lambda(q) && =  - 2q a_0 - \ln ( 1- 4 q^2)
\nonumber \\
 \lambda'(q) && =  - 2 a_0 + \frac{8q}{ 1- 4 q^2}
\label{lambdabox}
\end{eqnarray}

The second equation of the system \ref{legendrereciproque} 
\begin{eqnarray}
0 && =  2a+ \lambda'(q_a)   =  2 (a-a_0)  + \frac{8q_a}{ 1- 4 q_a^2}
\label{eq2box}
\end{eqnarray}
leads to the following second-order equation for $q_a$
\begin{eqnarray}
0 && =  q_a^2    - \frac{q_a}{ a-a_0}  - \frac{1}{4} 
\label{eqqa}
\end{eqnarray}
The appropriate solution $q_a$ that tends to $ q_a \to 0$ when $a \to a_0$ reads
\begin{eqnarray}
q_a = \frac{a_0-a}{2(  1+\sqrt{1+(a_0-a)^2} )} 
\label{soluqa}
\end{eqnarray}

The rate function given by the first equation of the system \ref{legendrereciproque}  reads
\begin{eqnarray}
I(a)  && = - \lambda(q_a)   -  2  a q_a =    2q_a (a_0-a) + \ln ( 1- 4 q_a^2)
 =   2q_a (a_0-a) + \ln (\frac{ 4 q_a}{ a_0-a} )
\nonumber \\
&& =\frac{(a_0-a)^2 }{ 1+\sqrt{1+(a_0-a)^2} } - \ln \left(  \frac{  1+\sqrt{1+(a_0-a)^2} }{2} \right)
\label{ratebox}
\end{eqnarray}

\section{ Statistical properties of the renormalized couplings }

\label{sec_jr}

In this section, we focus on the absolute values of the renormalized couplings given by Eq \ref{jrlow}
\begin{eqnarray}
\vert  J_{i_1,..,i_{r+1}}^{R^r}  \vert && = \vert  J_{i_1,...,i_{r+1}} \vert  \frac{P_{i_1,..,i_r} }{\sqrt{ 1+P^2_{i_1,..,i_r} }}
\label{jrlows}
\end{eqnarray}

\subsection{ Location of the critical point  }

On the Cayley tree where the number of points at distance $r$ grows exponentially as $K^r$,
the number of products $P(r)$ displaying the decay $P(r)\propto e^{-a r}$ reads (Eq. \ref{largedev})
\begin{eqnarray}
{\cal N} ( P(r) \propto e^{-a r}  ) && \oppropto_{r \to +\infty} K^r e^{-r I(a)} = e^{r (\ln K - I(a) )} \theta(a_{min} \leq a \leq a_{max})
\label{largedevN}
\end{eqnarray}
where the minimum value $a_{min}$ and the maximal value $a_{max}$ are respectively smaller and bigger than the typical value 
$a_{min}<a_0<a_{max}$ and satisfy
\begin{eqnarray}
I(a_{min})  = \ln K = I(a_{max})
\label{amin}
\end{eqnarray}
so that they occur only on a finite number $O(1)$ of branches,
while the typical value $a_0$ where $I(a_0)=0$ occur on an extensive $O(K^n)$ number of branches.

From Eq \ref{jrlows}, it is clear that the renormalized coupling $J(r)$ inherits the exponential decay of $P(r) $ of Eq. \ref{largedevN}
as long as $a>0$, while the region $a \leq 0$ produces  finite renormalized couplings $O(1)$ 
 so that the critical point corresponds to the vanishing of the minimal value $a_{min}$
\begin{eqnarray}
a_{min}^{criti}=0
\label{amincriti}
\end{eqnarray}
or equivalently in terms of the large deviation function $I(a)$ ( Eq. \ref{amin})
\begin{eqnarray}
I^{criti}(0)  = \ln K 
\label{criti}
\end{eqnarray}
For the special case of the box distribution of Eq \ref{box}, Eq \ref{ratebox}
yields the following explicit condition
in terms of the control parameter $a_0= \ln \frac{h}{J}$
\begin{eqnarray}
0 && =  \ln \left( K \frac{  1+\sqrt{1+(a^{criti}_0)^2} }{2} \right)- \frac{(a_0^{criti})^2 }{ 1+\sqrt{1+(a_0^{criti})^2} }
\label{critibox}
\end{eqnarray}

\subsection{ Paramagnetic phase for $a_{min}>0$  }

In the paramagnetic phase $a_{min}>0$, all $K^r$ renormalized couplings decay exponentially with $a \geq a_{min}>0$ (Eq \ref{largedevN})
\begin{eqnarray}
{\cal N} ( J(r) \propto e^{-a r}  ) && \oppropto_{r \to +\infty}  e^{r (\ln K - I(a) )} \theta(a_{min} \leq a \leq a_{max})
\label{largedevNj}
\end{eqnarray}

\subsection{ Spin-Glass phase for $a_{min}<0$  }

In the spin-glass phase $a_{min}<0$, the $K^r$ renormalized couplings can be split into two groups :
the number of finite couplings grows exponentially in $r$ as
\begin{eqnarray}
{\cal N} ( J(r) \propto O(1) ) && \oppropto_{r \to +\infty}  \int_{a_{min}}^0 da e^{r (\ln K - I(a) )} \simeq e^{r (\ln K - I(0) )} 
= e^{r (I(a_{min}) - I(0) )} 
\label{largedevNjfinite}
\end{eqnarray}
while the other branches are still characterized by exponential decays with exponents $a>0$
\begin{eqnarray}
{\cal N} ( J(r) \propto e^{-a r}  ) && \oppropto_{r \to +\infty}  e^{r (\ln K - I(a) )} \theta(0<  a \leq a_{max})
\label{largedevNjp}
\end{eqnarray}
This is the first indication that the ordered spin-glass cluster remains very sparse near the critical point,
as confirmed by the analysis of the renormalized transverse field in the next section.

\section{ Statistical properties of the renormalized transverse field }

\label{sec_hr}

In this section, we focus on the RG flow of Eq. \ref{hrlow} for the renormalized transverse field
\begin{eqnarray}
\ln \left( \frac{h_0^{R^r} }{ h_0^{R^{r-1}} } \right) 
&& \simeq  - \frac{1}{2}      \sum_{i_1=1}^{K+1} \sum_{i_2=1}^{K}  .. \sum_{i_r=1}^{K}
\ln \left( 1+ P_{i_1,..,i_{r-1},i_r}^2   \right) 
\label{hrlowsec}
\end{eqnarray}
which can be evaluated in terms of the large deviation analysis of Eq. \ref{largedevN}
 concerning the $K^r$ products $P(r)$
\begin{eqnarray}
\ln \left( \frac{h_0^{R^r} }{ h_0^{R^{r-1}} } \right) 
&& \opsimeq_{r \to +\infty}   - \frac{1}{2}   \int_{a_{min}}^{a_{max} } da  e^{r (\ln K - I(a) )} 
\ln \left( 1+ e^{-2ar}   \right) 
\label{hrlowlargedev}
\end{eqnarray}

\subsection{ Paramagnetic phase for $a_{min}>0$  }

In the paramagnetic phase $a_{min}>0$, Eq. \ref{hrlowlargedev} becomes
\begin{eqnarray}
\ln \left( \frac{h_0^{R^r} }{ h_0^{R^{r-1}} } \right) 
&& \opsimeq_{r \to +\infty}   - \frac{1}{2}   \int_{a_{min}}^{a_{max} } da  e^{r (\ln K - I(a) -2a  )}
= - \frac{1}{2}   \int_{a_{min}}^{a_{max} } da  e^{r (I(a_{min}) - I(a) -2a  )} 
\label{hrlowlargedevpara}
\end{eqnarray}
The integral is dominated by the lower boundary $a_{min}$ of the integral, and one obtains the exponential decay
\begin{eqnarray}
\ln \left( \frac{h_0^{R^r} }{ h_0^{R^{r-1}} } \right) 
&& \oppropto_{r \to +\infty}   - e^{-2 a_{min} r } 
\label{hrlowlargedevparaexp}
\end{eqnarray}
By integration, one obtains that $h_0^{R^r} $ remains finite as $r \to +\infty$
\begin{eqnarray}
\ln \left( \frac{h_0^{R^r} }{ h_0 } \right) 
&&  \oppropto_{r \to +\infty} -    \int_{1}^{r} dr'  e^{ -2a_{min}  r'} \oppropto_{r \to +\infty}   - \frac{1-e^{-2 a_{min} r }  }{ a_{min} }  
\label{hrlowlargedevparafinite}
\end{eqnarray}
The typical asymptotic value $h_0^{R^{\infty}} $ for the renormalized transverse field
diverges with the following essential singularity near the transition $a_{min} \to a_{min}^{criti}=0$
\begin{eqnarray}
\ln \left( \frac{h_0^{R^{\infty}} }{ h_0 } \right)  &&  \oppropto_{a_{min} \to 0}   - \frac{1}{ a_{min} }  
\label{essential}
\end{eqnarray}

\subsection{ Spin-Glass phase for $a_{min}<0$  }

In the spin-glass phase $a_{min}<0$, 
it is convenient to evaluate separately the contributions of the two regions $a<0$ and $a>0$ in 
the integral of Eq. \ref{hrlowlargedev}.
The contribution of the region $a>0$ is dominated by the lower boundary $a=0$ of the integral
\begin{eqnarray}
  \int_{0}^{a_{max} } da  e^{r (\ln K - I(a) )}  \ln \left( 1+ e^{-2ar}   \right) 
\opsimeq_{r \to +\infty}   \int_{0}^{a_{max} } da  e^{r (\ln K - I(a)-2a )}  \simeq  e^{r (\ln K - I(0) )} = e^{r (I(a_{min}) - I(0) )}
\label{hrlowlargedevp}
\end{eqnarray}
corresponding to an exponentially growing term.
The region $a<0$
\begin{eqnarray}
  \int_{a_{min}}^{0 } da  e^{r (\ln K - I(a) )}  \ln \left( 1+ e^{-2ar}   \right) 
\opsimeq_{r \to +\infty}  \int_{a_{min}}^{0 } da  e^{r (\ln K - I(a) )}  (-2ar)
\label{hrlowlargedevm}
\end{eqnarray}
is dominated by the upper boundary $a=0$.
So the RG flow of renormalized transverse field of Eq. \ref{hrlowlargedev} is dominated by the exponentially big term
of coefficient $(I(a_{min}) - I(0))>0$ of Eq. \ref{hrlowlargedevp}
\begin{eqnarray}
\ln \left( \frac{h_0^{R^r} }{ h_0 } \right) 
&& \oppropto_{r \to +\infty}
  -   \int_{1}^{r} dr'  e^{r' (I(a_{min}) - I(0) )}  \oppropto_{r \to +\infty}   - \frac{ e^{r (I(a_{min}) - I(0) )} }{(I(a_{min}) - I(0) )}
\label{hrlowlargedevsg}
\end{eqnarray}

\subsection{ Finite-size scaling in the critical region  }

The above results for the renormalized transverse field as a function of the radial distance $r$ can be summarized by
the following finite-size scaling form in the critical region
\begin{eqnarray}
\ln \left( \frac{h_0^{R^r} }{ h_0 } \right) 
&& \oppropto_{r \to +\infty}
  -   r^{\psi} G \left( r^{\frac{1}{\nu}} (J-J_c) \right) 
\label{hrlowfss}
\end{eqnarray}
with the exponent 
\begin{eqnarray}
\psi=1
\label{ps1}
\end{eqnarray}
and the correlation length exponent
\begin{eqnarray}
\nu=1
\label{nu1}
\end{eqnarray}
as in many other phase transitions on the Cayley tree.
The scaling function $G(x)$ is constant at the origin $G(0)=cst$, behaves as
\begin{eqnarray}
G(x) \oppropto_{x \to - \infty} - \frac{1}{x} 
\label{gminfty}
\end{eqnarray}
to reproduce the behavior of Eq. \ref{essential} in the paramagnetic phase $J<J_c$, and as
\begin{eqnarray}
G(x) \oppropto_{x \to + \infty}  \frac{e^x-1}{x} 
\label{gpinfty}
\end{eqnarray}
to reproduce the behavior of Eq. \ref{hrlowlargedevsg} in the spin-glass phase $J>J_c$.

\subsection{ Number $N_{SG}$ of spins involved in this ordered spin-glass cluster }

The renormalized transverse field $h_0^{R^r} $ directly reflects the number $N_{SG}(r)$ of spins involved in this ordered spin-glass cluster
\begin{eqnarray}
\ln \left( \frac{h_0^{R^r} }{ h_0 } \right) 
&& \oppropto_{r \to +\infty}  -  N_{SG}(r)
\label{hnsg}
\end{eqnarray}
In the paramagnetic phase, both remain finite as $r \to +\infty$.
In the spin-glass phase,
The behavior found in Eq. \ref{hrlowlargedevsg}
 for the renormalized transverse field
thus confirms the indication of Eqs \ref{largedevNjfinite} and \ref{largedevNjp}
concerning the renormalized couplings : near the critical point, the 
ordered spin-glass cluster remains very sparse. 
More precisely, the number $N_{SG}$ of spins involved in this ordered spin-glass cluster grows exponentially with the distance $r$
\begin{eqnarray}
N_{SG} \propto  e^{r (I(a_{min}) - I(0) )}  = e^{r (\ln K - I(0) )} 
\label{nsg}
\end{eqnarray}
but is only sub-extensive with respect to the total number of spins $N=K^r$
\begin{eqnarray}
N_{SG} \propto  = e^{r (\ln K - I(0) )} =N^{\theta}
\label{nsgn}
\end{eqnarray}
in the whole region of the phase diagram where the continuously varying exponent
\begin{eqnarray}
\theta =  1- \frac{I(0)}{\ln K } = 1 - \frac{I(0)}{I(a_{min}) } 
\label{theta}
\end{eqnarray}
remains in the interval
\begin{eqnarray}
\theta^{criti}=0 <\theta <  1 = \theta^{ext}
\label{thetac}
\end{eqnarray}
At criticality, the vanishing exponent $\theta^{criti}=0 $ corresponds to the logarithmic growth with respect to $N$ (Eq \ref{hrlowfss} and \ref{ps1})
\begin{eqnarray}
N_{SG}^{criti} \propto r  = \frac{ \ln N }{ \ln K}
\label{nsgnln}
\end{eqnarray}
meaning that only a finite number of the branches sustain the spin-glass order.
The location where the spin-glass-ordered cluster becomes extensive $\theta^{ext}=1$ corresponds to 
the vanishing of the large deviation rate function $I^{ext}(0)=0$, i.e. to the vanishing of the typical value $a_0^{ext}=0$
(Eq. \ref{largedevtyp}), i.e. to the location of the transition for the one-dimensional chain (Eq \ref{criti1d})
\begin{eqnarray}
a_0^{ext}=0= a_0^{criti1d}
\label{a0ext}
\end{eqnarray}

The finite region of the phase diagram corresponding to Eq. \ref{thetac} where the ordered spin-glass cluster remains sub-extensive 
is somewhat { \it formally } reminiscent of the delocalized non-ergodic phase
 existing in the Anderson Localization model defined on the Cayley tree
\cite{levitov,us_cayley,mirlin_tik}, i.e. in exactly the same geometry as in the present paper, and for the same technical reasons
based on large deviations on the branches of the Cayley tree \cite{levitov}.
But of course the physical meaning of the phases is completely different : in the Anderson Localization Model,
the three phases are Localized/Non-Ergodic-Delocalized/Ergodic-Delocalized,
while in the present study, the three phases are all MB-Localized, namely 
Paramagnetic-MBL/SG-MBL with sub-extensive SG-cluster/SG-MBL with extensive SG-cluster.

Let us mention however that the existence of the intermediate delocalized non-ergodic phase remains very controversial 
for the Anderson Localization model on Random Regular Graphs 
\cite{biroli_nonergo,luca,mirlin_ergo,altshuler_nonergo,lemarie,ioffe}
or for Many-Body-Localization models \cite{grover,harrisMBL,c_entropy,c_mblrgeigen,santos},
where an analogy with the Anderson Localization transition in an Hilbert space of 'infinite dimensionality'
 has been put forward \cite{levitov,gornyi_fock,vadim,us_mblaoki,gornyi},
while the properties of the delocalized non-ergodic phase can be explicitly computed in some random matrix models
\cite{kravtsov_rosen,biroli_rosen,ossipov_rosen,c_ww}.
For our present study, these results thus indicate that the intermediate SpinGlass-MBLocalized phase with sub-extensive SG-cluster
found here on the Cayley tree might not exist on other tree-like lattices like Random Regular Graphs.

\subsection { Physical meaning of the results }

\begin{figure}
\begin{center}
\includegraphics[width=16cm]{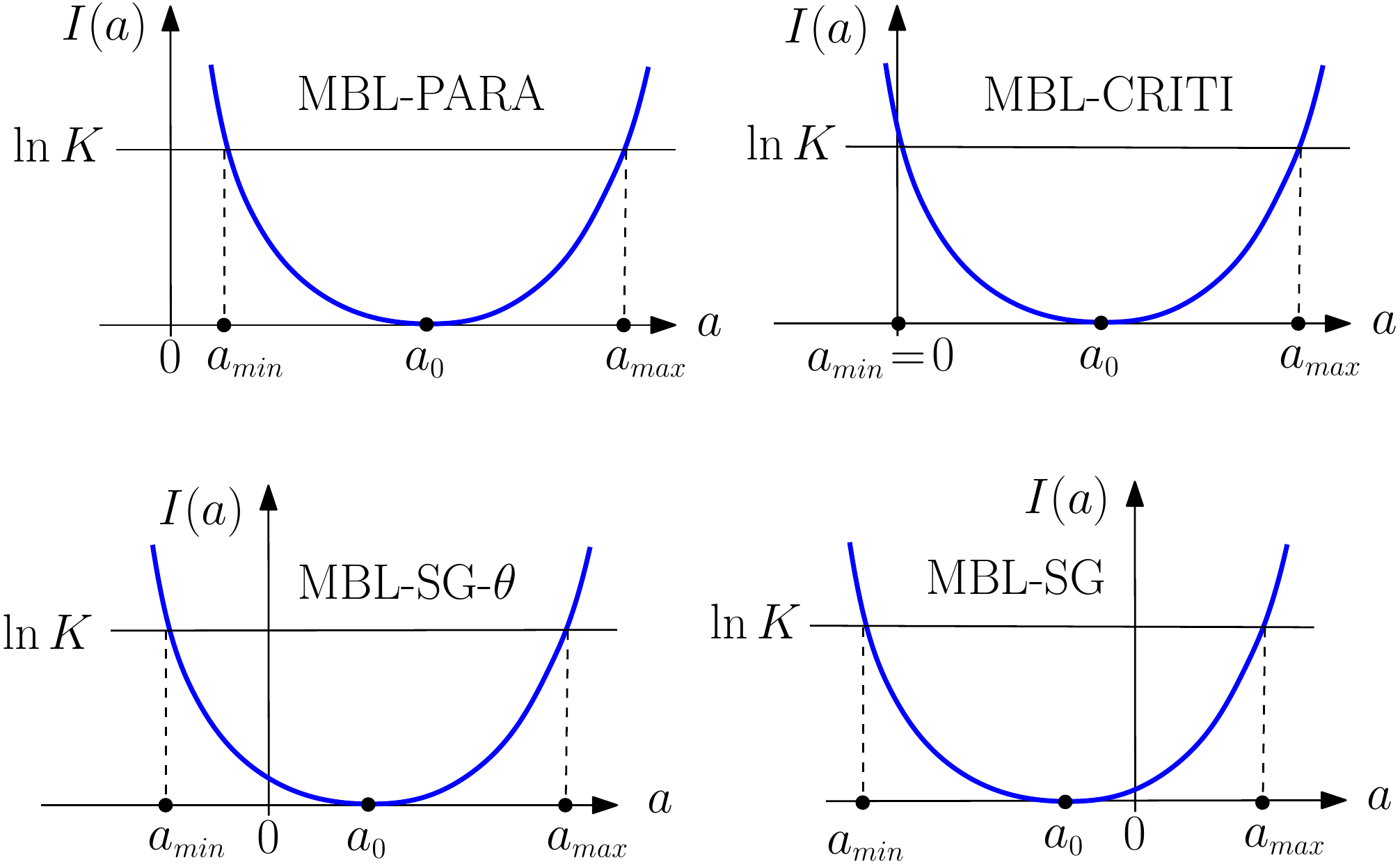}
\end{center}
\caption{ The two important control parameters for the phase diagram are the minimal value $a_{min}$ and the typical value $a_0 = \overline{  (\ln  \vert  h_{i} \vert - \ln \vert J_{i}   \vert ) } $ where the large deviation function $I(a)$ takes respectively the values $I(a_{min})= \ln K$ and $I(a_0)=0$. The critical point $a_{min}=0$ corresponds to the phase transition between the MBL-Paramagnetic phase $a_{min}>0$ and the MBL-Spin-Glass phase $a_{min}<0$, where the Spin-Glass-Order remains sub-extensive $0<\theta=1- \frac{I(0)}{\ln K }  <1$ in the whole region $a_{min}<0<a_0$ before becoming extensive  in the region $a_0<0$.}
\label{figure}
\end{figure}

The above results can be summarized as follows (see Figure \ref{figure}).

\subsubsection{ The two important control parameters }

The two important control parameters for the quantum Ising model on the Cayley tree of branching ratio $K$ are

(1) the typical value of Eq. \ref{a0} where the large deviation function $I(a)$ of Eq. \ref{largedev} vanishes $I(a_0)=0$
\begin{eqnarray}
a_{0} \equiv \overline{  (\ln  \vert  h_{i} \vert - \ln \vert J_{i}   \vert ) } 
\label{a0def}
\end{eqnarray}

(2) the minimum value $a_{min}$ defined as the smaller value $a_{min}<a_0$ where the large deviation function $I(a)$ of Eq. \ref{largedev}
takes the value
\begin{eqnarray}
I(a_{min})  = \ln K 
\label{amindef}
\end{eqnarray}

\subsubsection{ The three possible MB-Localized phases }

(a) {\it MB-Localized Phase with extensive Spin-Glass Order }

 In the region where the typical value $a_0$ of Eq. \ref{a0} is negative
\begin{eqnarray}
a_{0} <0
\label{a0order}
\end{eqnarray}
a typical one-dimensional chain would be spin-glass ordered (Eq. \ref{criti1d}), and thus the whole Cayley tree is also fully ordered
with an extensive spin-glass cluster with respect to the total number of spins $N=K^r$
\begin{eqnarray}
N_{SG} \propto   K^r=N
\label{nsgorder}
\end{eqnarray}

(b) {\it Paramagnetic MB-Localized Phase }

In the region where the minimal value $a_{min}$ is positive
\begin{eqnarray}
0< a_{min} 
\label{aminpara}
\end{eqnarray}
the drawing of $K^r$ independent random one-dimensional chains would produce only paramagnetic chains,
i.e. even the exponentially-rare best chain would be paramagnetic. Then the whole Cayley tree is also paramagnetic,
and the spin-glass cluster around the origin remains finite
\begin{eqnarray}
N_{SG} \propto  O(1)
\label{nsgpara}
\end{eqnarray}

(c) {\it MB-Localized Phase with sub-extensive Spin-Glass Order }

In the intermediate region
\begin{eqnarray}
 a_{min} <0 < a_0
\label{amininter}
\end{eqnarray}
corresponding to
\begin{eqnarray}
 I(a_{min})= \ln K  > I(0)  > I( a_0 )=0
\label{iamininter}
\end{eqnarray}
the drawing of $K^r$ independent random one-dimensional chains would produce 
$K^r  e^{- r I(0) )} $ spin-glass ordered chains, while the other (of order $K^r$) would be paramagnetic.
Then on the Cayley tree, the spin-glass cluster around the origin only contains $K^r  e^{- r I(0) )} $ leaves out of the $K^r$.
So the size of the spin-glass cluster grows exponentially in $r$ but not as rapidly as $N=K^r$, so that it is subextensive
\begin{eqnarray}
N_{SG} \propto  = e^{r (\ln K - I(0) )} =N^{\theta}
\label{nsgninter}
\end{eqnarray}
 where the exponent
\begin{eqnarray}
\theta =  1- \frac{I(0)}{\ln K } = 1 - \frac{I(0)}{I(a_{min}) } 
\label{thetabis}
\end{eqnarray}
varies continuously between $\theta^{criti}=0 $ [corresponding to $a_{min}=0$ where the transition towards (b) occurs]
and $\theta^{ext}=1 $ [corresponding to $I(0)=0$ i.e. $a_0=0$ where the transition towards (a) occurs].

\section{Conclusion }

\label{sec_conclusion}

We have introduced a simple Real-Space-Renormalization procedure in order to construct
the whole set of eigenstates for
the quantum Ising model with random couplings and random transverse fields on the Cayley tree of branching ratio $K$.
 The analysis of the renormalization rules via large deviations was described to obtain the critical properties
of  the phase transition between the paramagnetic and the spin-glass Many-Body-Localized phases.
 In particular, we have found that the renormalized transverse field of the center site
involves the activated exponent $\psi=1$ and the correlation length exponent $\nu=1$. 
The spin-glass-ordered cluster containing $N_{SG}$ spins was found to be extremely sparse with respect to the total number $N \propto K^r$ of spins : its size grows only logarithmically at the critical point $N_{SG}^{criti} \propto \ln N$, meaning that only a finite number $O(1)$ of the branches are long-ranged-ordered, while the other branches display exponentially decaying correlations.
In addition, the size $N_{SG} $ spin-glass-ordered cluster is sub-extensive $N_{SG} \propto N^{\theta}$ in the finite region of the spin-glass phase where the continuously varying exponent $\theta$ remains in the interval $0<\theta<1$.

As a final remark, let us mention that the mere existence of Many-Body-Localized phases in any dimension $d>1$ has been recently challenged 
\cite{chandran-dg1,roeck-dg1,imbrie-dg1}, the same arguments being also used to claim the impossibility of mobility edges for MBL in $d=1$ \cite{bubbles}
(as opposed to the numerical phase-diagrams found in Ref. \cite{kjall,alet,luitz_tail,mondragon})
as well as the impossibility of MBL in the presence of power-law interactions \cite{roeck-dg1}
 (as opposed to the works \cite{yao,burin_fss,burin_xy,hauke,gutman,moessner,hi,sondhi}).
It is thus essential to study various MBL models in various dimensions $d>1$
in order to solve the controversial issue about the influence of the dimension $d$.
 Many-Body-Localized phases have been reported in dimension $d=2$ both numerically \cite{regnault2d} and experimentally \cite{mbl2d_exp},
as well as on Random Regular graphs \cite{mossi} or in the mean-field quantum random energy model \cite{qrem1,qrem2}.
We thus hope that the present work concerning Many-Body-Localized phases on the Cayley tree of effective infinite dimension $d=\infty$
will motivate future studies on this topic.

\end{document}